\documentstyle[12pt,iopconf]{article}
\newcommand{\de}{\partial}
\newcommand{\be}{\begin{equation}}
\newcommand{\ee}{\end{equation}}
\newcommand{\bea}{\begin{eqnarray}}
\newcommand{\eea}{\end{eqnarray}}
\def\nn{\nonumber}
\def\f{\varphi}
\def\th{\theta}
\def\p{\partial}
\def\a{\alpha}
\def\b{\beta}
\def\n{\nabla}
\def\o{\omega}
\def\O{\Omega}
\def\g{\gamma}
\def\s{\sigma}

\begin{document}
\title{Detection of Scalar Gravitational Waves}

\author{Francesco Fucito\footnote{E-mail: Fucito@roma2.infn.it.}}

\affil{INFN, sez. di Roma 2, Via della Ricerca Scientifica, 00133 Rome}

\beginabstract
In this talk I review recent progresses in the detection 
of scalar gravitational waves. Furthermore, in the framework of
the Jordan-Brans-Dicke theory, I compute the signal to noise ratio
for a resonant mass detector of spherical shape and for 
binary sources and collapsing stars. Finally I compare these results
with those obtained from laser interferometers and from 
Einsteinian gravity.
 
\endabstract

\section{Introduction}
The efforts aimed at the detection of  gravitational waves (GW) 
started more than a quarter of century ago and have been,
so far, unsuccessful \cite{ama, odia}. Resonant bars 
have proved their reliability, being capable of continous
data gathering for long periods of time \cite{asto, joal}. 
Their energy sensitivity
has improved of more than four orders of magnitude since Weber's
pioneering experiment. But a  further improvement is still 
necessary to
achieve successful detection. While further developments of 
bar detectors are under way, two new generations of earth 
based experiments
have been proposed: detectors based on large laser 
interferometers are already under construction \cite{abbra}, 
resonant detectors of spherical shape are under 
study \cite{odia}.

In this lecture I report on a series of papers \cite{fuc} in which 
the opportunity of introducing resonant mass detectors of spherical
shape was studied. As a general motivation for their study, spherical
detectors have the advantage over bar-shaped detectors of a larger
degree of symmetry. This translates into the possibility of building
detectors of greater mass and consequently of higher cross section.

Besides this obvious observation, the higher degree 
of symmetry enjoyed
by the spherical shape puts such a detector in the unique position 
of being able of detecting GW's with a spin content different from 2.
This means to test non-Einsteinian theories of gravity.

I would in fact now like to remind the 
reader of the very special position of Einstein's general 
relativity (GR)
among the possible gravitational theories. 
Theories of gravitation, in fact,
can be divided into two families: metric and non-metric 
theories \cite{will}.
The former can be defined to be all theories obeying the following
three postulates:
\begin{itemize}
\item
spacetime is endowed with a metric; 
\item
the world lines of test particles are geodesic of the 
above mentioned
metric;
\item
in local free-falling frames, the non-gravitational 
laws of physics 
are those 
of special relativity.
\end{itemize}
It is an obvious consequence of these postulates 
that a metric 
theory obeys the principle  of equivalence. More 
succintly a theory 
is said 
to be metric if the action of gravitation on the matter sector 
is due exclusively to the metric tensor.
GR is the most famous example of a metric 
theory. Kaluza-Klein type theories, also belong to this
class along with the Brans-Dicke theory. Different 
representatives of 
this class differ for their equations of motion 
which in turn can be 
deduced 
from a lagrangian principle. 
Since there seems to be no compelling experimental
or theoretical reasons to introduce non-Einsteinian
or non-metric theories, they are sometimes considered a 
curiosity.
This point should perhaps be reconsidered. String theories are 
in fact the most serious candidate for a theory of quantum gravity,
the standard cosmological model has been emended with 
the introduction of inflation and even the introduction of 
a cosmological constant (which seems to be needed to explain recent
cosmological data) could imply the  existence of 
other gravitationally coupled fields. In all of the above
cited cases I am forced to introduce fields which 
are non-metrically coupled in the sense explained above.

In the first section of this lecture I will explain that a spherical
detector is able to detect any spin component of an impinging GW.
Moreover its vibrational eigenvalues can be divided into two sets 
called spheroidal and toroidal. Only the first set couples to the
metric. This leads to the opportunity of using such a detector as
a veto for non-Einsteinian theories. In the second section I take
as a model the Jordan-Brans-Dicke (JBD), in which along with 
the metric I also have a scalar field which is metrically coupled.
I am then able to study the signal to noise ratio for sources such
as binary systems and collapsing stars and compare the strenght
of the scalar signal with respect to the tensor one. Finally in the
third and last section I repeat this computation in the case of the
hollow sphere which seems to be the detector which is most likely 
to be built.

\section{Testing Theories of Gravity\label{sec:grav}}
\setcounter{equation}{0}
\subsection{Free Vibrations of an Elastic Sphere}
Before discussing the interaction with an external GW field,
let us consider the basic equations governing the free vibrations
of a perfectly homogeneous, isotropic sphere of radius $R$, 
made of a material having density $\rho$ and Lam\'e coefficients 
$\lambda$ 
and $\mu$ \cite{elast}. 
 	
Following the notation of \cite{lobo}, let $x_{i}, i=1,2,3$ be 
the equilibrium position of the element of the
elastic sphere and $x'_{i}$ be the deformed position then
$u_{i}=x'_{i}-x_{i}$ is the displacement vector. Such vector 
is assumed
small, so that the linear theory of elasticity is applicable. 
The strain 
tensor is defined as $u_{ij}=(1/2)(u_{i,j}+u_{j,i})$
and is related to the stress tensor by
$\sigma_{ij}=\delta_{ij}\lambda u_{ll}+2\mu u_{ij}$.
The equations of motion of the free vibrating sphere are thus
\cite{elast}
\be
\rho \frac {\de ^{2}u_{i}}{\de  t^{2}}= \frac {\de }{\de 
 x^{j}}(\delta_{ij}\lambda u_{ll}+2\mu u_{ij}) 
\ee
with the boundary condition:
\be
n_{j}\sigma _{ij}=0
\label{contorno}
\ee
at $r=R$ where $n_i\equiv x_{i}/r$ is the unit normal. 
These conditions simply state that the surface of the sphere is 
free to vibrate.
The displacement $u_{i}$ is a time-dependent vector, whose time 
dependence can be factorised as 
$u_{i}({\vec x},t)=u_{i}({\vec x})exp(i\omega t)$,
where $\omega$ is the frequency. The equations of motion 
then become:
\be
\mu \nabla ^{2}{u}_i+(\lambda +\mu )\nabla_i 
(\nabla_j u_j)=-\omega 
^{2}\rho {u}_i 
\label{eqmotomio}
\ee
Their solutions can be expressed as a sum of a longitudinal and 
two transverse 
vectors \cite{drei}:
\be 
\vec{u}(\vec{x})=C_{0}\vec{\nabla}\phi (\vec{x})+C_{1}
\vec{L}\chi (\vec{x})+C_{2}\vec\nabla\times\vec{L}\chi (\vec{x}) 
\label{usol}
\ee
where $C_{0}, C_{1}, C_{2}$ are dimensioned constants and $\vec{L}
\equiv\vec{x}\times
\vec{\nabla}$ is the angular momentum operator.
Regularity at $r=0$ restricts the scalar functions 
$\phi$ and $\chi$  
to be
expressed as $\phi(r,\theta,\varphi)\equiv j_{l}(qr)Y_{lm}
(\theta ,\varphi)$ and
$\chi(r,\theta,\varphi)\equiv j_{l}(kr)Y_{lm}(\theta ,\varphi)$. 
$Y_{lm}(\theta ,
\varphi)$ are the spherical harmonics and $j_{l}$ the 
spherical Bessel 
functions \cite{jack}: 
\be
j_l(x)=\bigg({1\over x}{d\over dx}\bigg)^l
\bigg(\frac{\sin x}{x}\bigg)
\label{bessel}
\ee
$q^{2}\equiv\rho\omega ^{2}/(\lambda +2\mu)$ and 
$k^{2}\equiv\rho\omega ^{2}/\mu$ are the longitudinal 
and transverse 
wave vectors respectively.

Imposing the boundary conditions (\ref{contorno}) 
at $r=R$ yields 
two families 
of solutions:
\begin{itemize}
\item
{\em Toroidal} modes: these are obtained by setting 
$C_{0}=C_{2}=0$, and $C_{1}
\neq 0$. In this case the displacements in (\ref{usol})
can be written in terms of
the basis:
\be
\vec{\psi}^{T}_{nlm}(r,\theta,\varphi)=T_{nl}(r)\vec{L}Y_{lm}
(\theta,\varphi)
\label{paperone} 
\ee
with $T_{nl}(r)$ proportional to $j_{l}(k_{nl}r)$.
The eigenfrequencies are
determined by the boundary conditions (\ref{contorno}) which read
\cite{drei}
\be
f_{1}(kR)=0
\ee
where 
\be
f_{1}(z)\equiv\frac{d}{dz}\bigg[\frac{j_{l}(z)}{z}\bigg]. 
\ee

\item

{\em Spheroidal} modes: these are obtained by 
setting $C_{1}=0$, $C_{0}\neq 0$ and $C_{2}\neq 0$. 
The displacements of (\ref{usol}) can be expanded in the
basis
\be
\vec{\psi}^{S}_{nlm}(\vec{x})=A_{nl}(r)Y_{lm}(\theta ,\varphi)\vec{n}
-B_{nl}(r)\vec{n} \times\vec{L}Y_{lm}(\theta ,\varphi)
\label{quattrostagioni}
\ee
where $A_{nl}(r)$ and $B_{nl}(r)$ are dimensionless radial 
eigenfunctions 
\cite{lobo}, which can be expressed in terms of the
spherical Bessel functions and their derivatives. 
The eigenfrequencies are determined by the boundary conditions 
(\ref{contorno}) which read \cite{lobo}
\be
det\pmatrix{f_{2}(qR)-{\lambda\over 2\mu}q^{2}R^{2}f_{0}(qR)&l(l+1)
f_{1}(kR)\cr f_{1}(qR)&{1\over 2}f_{2}(kR)+[\frac{l(l+1)}{2}-1]
f_{0}(kR)}=0
\label{minnie}
\ee
where 
\be
f_{0}(z)\equiv\frac{j_{l}(z)}{z^{2}} \quad
f_{2}(z)\equiv\frac{d^{2}}{dz^{2}}j_{l}(z)
\ee
\end{itemize}
The eigenfrequencies can be determined numerically for both 
toroidal and 
spheroidal vibrations. Each mode of order $l$ is $(2l+1)$-fold 
degenerate.
The eigenfrequency values can be obtained from :
\be
\omega_{nl} = \sqrt{\mu\over\rho} {(kR)_{nl}\over R}
\ee
\subsection{Interaction of a Metric GW with the Sphere Vibrational 
Modes}
The detector is assumed to be non-relativistic 
(with sound velocity $v_s\ll c$ and radius $R\ll \lambda$
the GW wavelength) and 
endowed with a high quality factor ($Q_{nl}=\omega_{nl} \tau_{nl}\gg1$, 
where $\tau_{nl}$ is the decay time of the mode $nl$). 
The displacement $\vec u$ of a point in the detector can be 
decomposed in 
normal modes as:
\be
\vec u(\vec{x},t) = \sum_N A_N(t) \vec \psi_N (\vec x)
\label{expan}
\ee
where $N$ collectively denotes the set of quantum numbers 
identifying the mode.
The basic equation governing the response of the detector is 
\cite{mtw}
\be
\ddot{A}_{N}(t) + \tau _{N}^{-1} \dot{A}_{N}(t) + 
\omega _{N}^{2}A_{N}(t) = 
f_{N}(t)
\label{forzate}
\ee
I assume that the gravitational interaction obeys the 
principle of equivalence 
which has been experimentally supported to high accuracy.
In terms of the 
so-called electric components of the Riemann tensor 
$E_{ij}\equiv R_{0i0j}$,
the driving force $f_N(t)$ is then given by \cite{wago}
\be
f_{N}(t) = - M^{-1} E_{ij}(t) \int \psi_{N}^{i*}(\vec{x}) x^{j}
\rho d^{3}x
\label{forza}
\ee
where $M$ is the sphere mass and I consider the density 
$\rho$ as a constant.
In any metric theory of gravity $E_{ij}$ is a $3\times3$
symmetric tensor, which depends on time, but not on 
spatial coordinates.

Let us now investigate 
which sphere 
eigenmodes can be excited by a metric GW, {\it i.e.} which 
sets of quantum 
numbers N give a non-zero driving force.

a) {\em Toroidal} modes 

\noindent The eigenmode vector, $\psi^T_{nlm}$ can be expressed 
as in eq. (\ref{paperone}).
Up to an adimensional normalisation constant C, the driving force is
\bea
f^{(T)}_{N}(t)&=&-e^{-i\omega_{N}t}\frac{3C}{4\pi 
R^{3}}\int_{0}^{R}dr r^{3}
j_{l}(k_{nl}^{(T)}r)\int_{0}^{\pi}d\theta
\sin\theta\int_{0}^{2\pi}d\phi
\nonumber\\
& &\bigg{\{}\frac{E_{yy}-E_{xx}}{2} \bigg(\sin\theta 
\sin 2\phi\frac{\de  Y_{lm}^{*}}
{\de \theta}+\cos\theta\cos 2\phi\frac{\de  Y_{lm}^{*}}
{\de \phi}\bigg)\nonumber \\
& & \mbox{}+E_{xy}\bigg(\sin\theta\cos 2\phi\frac{\de  
Y_{lm}^{*}}{\de \theta}-
\cos\theta\sin 2\phi\frac{\de  Y_{lm}^{*}}{\de \phi}\bigg)
\nonumber \\
& & \mbox{}+E_{xz}\bigg[-\sin\phi\cos\theta\frac{\de  
Y_{lm}^{*}}{\de \theta}+
(\sin\theta\cos\phi-\frac{\cos^2\theta}{\sin\theta}\cos\phi)
\frac{\de  Y_{lm}^{*}}{\de \phi}\bigg]\nonumber \\
& & \mbox{}+E_{yz}\bigg[\cos\phi\cos\theta\frac{\de 
Y_{lm}^{*}}{\de\theta}
+(\sin\theta\sin\phi-\frac{\cos^2\theta}{\sin\theta}\sin\phi) 
\frac{\de Y_{lm}^{*}}{\de \phi}\bigg]\nonumber \\
& & \mbox{}+\bigg(E_{zz}-\frac{E_{xx}+E_{yy}}{2}\bigg)
\cos\theta\frac{\de  Y_{lm}^{*}}{\de \phi}\bigg\}
\label{intforz}
\eea
Using the equations
\be
\frac{\de  Y_{lm}^{*}}{\de \theta}= (-)^m 
\bigg\lbrack\frac{2l+1}{4\pi}\frac{(l-m)!}{(l+m)!}
\bigg\rbrack^{1\over 2} 
\frac{\de  P_{l}^{m}(\cos\theta)}{\de \theta}e^{-im\phi}
\label{legendre}
\ee
and
\be
\frac{\de  Y_{lm}^{*}}{\de \phi}= -im(-)^m 
\bigg\lbrack\frac{2l+1}{4\pi}\frac{(l-m)!}{(l+m)!}
\bigg\rbrack^{1\over 2}
P_{l}^{m}(\cos\theta)e^{-im\phi}
\label{lavendetta}
\ee
the integration over $\phi$ can be performed. Eq. (\ref{intforz}) 
then contains
integrals over $\theta$ of the form:
\be
\int_0^\pi \bigg[(\sin^2\theta-\cos^2\theta)P_l^{\pm 1}(\cos\theta)
- \sin\theta\cos\theta\frac{\de P_l^{\pm 1}(\cos\theta)}
{\de \theta}\bigg]d\theta
\label{carolina}
\ee
and 
\be
\int_0^\pi \bigg[2\sin\theta\cos\theta P_l^{\pm 2}(\cos\theta)
+ \sin^2\theta\frac{\de P_l^{\pm 2}(\cos\theta)}{\de \theta}
\bigg]d\theta
\label{annamaria}
\ee
After integration by parts, the derivative terms in eqs. 
(\ref{carolina}) and 
(\ref{annamaria}) exactly cancel the non-derivative ones.
The remaining boundary terms vanish too, thanks to the 
periodicity of the trigonometric functions and to the 
regularity of the 
associated Legendre polynomials. 
The vanishing of the above integrals has a profound physical 
consequence.
It means that in any metric theory of gravity the toroidal modes of 
the sphere 
cannot be excited by GW and can thus be used as a veto in the 
detection. 
\par
b) {\em Spheroidal} modes \par 
\noindent The forcing term is given by:
\be
f^{(S)}_{N}(t) = - M^{-1} E_{ij}(t) \int x^{j}
\bigg({x^i\over r} A_N (r) 
Y_{lm}(\theta,\varphi) - B_N (r) \epsilon^{ink} {x_n\over r} L_k 
Y_{lm}(\theta,\varphi)\bigg) \rho d^{3}x
\label{sferforza}
\ee
One is thus lead to compute integrals of the following types
\be
\int x^{j}x^i Y_{lm}(\theta,\varphi) d^3 x
\ee
and
\be
\int x^{j}x^i L_k Y_{lm}(\theta,\varphi) d^3 x
\ee
Since the product $x^i x^j$ can be expressed in terms of the 
spherical
harmonics with $l=0,2$ and the angular momentum operator does 
not change the 
value of $l$, one immediately concludes that in any metric 
theory of gravity
only the $l=0,2$ spheroidal modes of the sphere can be excited. 
At the lowest level there are a total of
five plus one independent spheroidal modes that can be used for 
GW detection and study.

\subsection{Measurements of the Sphere Vibrations and Wave 
Polarization States}
From the analysis of the spheroidal modes active for 
metric GW, I now want
to infer the field content of the theory.
For this purpose it is convenient to express the Riemann tensor in a 
null (Newman-Penrose) tetrad basis \cite{will}.

To lowest non-trivial order in the perturbation the six indipendent 
"electric" components of the Riemann tensor may be expressed in terms
of the Newmann-Penrose (NP) parameters as 
\be
E_{ij} = \pmatrix{-Re\Psi_4-\Phi_{22}&Im\Psi_4&-2\sqrt2 Re\Psi_3\cr 
Im\Psi_4&Re\Psi_4-\Phi_{22}&2\sqrt2 Im\Psi_3\cr 
-2\sqrt2 Re\Psi_3&2\sqrt2 Im\Psi_3&-6\Psi_2}  
\label{newpen}
\ee 
The NP parameters allow the identification of the spin
content of the metric theory responsible for the generation of the wave
\cite{will}. The classification can be summarized in order of incresing
complexity as follows:
\begin{itemize}
\item
General Relativity (spin 2): $\Psi_4\neq 0$ while 
$\Psi_2=\Psi_3=\Phi_{22}=0$.
\item
Tensor-scalar theories (spin 2 and 0): $\Psi_4\neq 0$, $\Psi_3=0$,
$\Psi_2\neq 0$ and/or $\Phi_{22}\neq 0$ ({\it e.g.} Brans-Dicke theory
with $\Psi_4\neq 0$, $\Psi_2=0$, $\Psi_3=0$ and $\Phi_{22}\neq 0$).
\item
Tensor-vector theories (spin 2 and 1):
$\Psi_4\neq 0$, $\Psi_3\neq 0$, $\Phi_{22}=\Psi_2=0$.
\item
Most General Metric Theory (spin 2, 1 and 0): 
$\Psi_4\neq 0$, $\Psi_2\neq 0$, $\Psi_3\neq 0$ and $\Phi_{22}\neq 0$, 
({\it e.g.} Kaluza-Klein theories with $\Psi_4\neq 0$, $\Psi_3\neq 0$, 
$\Phi_{22}\neq 0$ while $\Psi_2=0$).

\end{itemize}

In eq. (\ref{newpen}), I have assumed that the wave comes from a 
localized 
source with wave vector $\vec k$ parallel to the $z$ axis of 
the detector 
frame. 
In this case the NP parameters (and thus the wave polarisation 
states) can be
uniquely determined by monitoring the six lowest spheroidal modes.
If the direction of the incoming wave is not known two 
more unknowns appear in the problem, {\it i.e.} 
the two angles of rotation of the 
detector frame needed to align $\vec k$ along the $z$ axis. 
In order to dispose of this problem one can envisage the 
possibility of 
combining the pieces of information from an array of 
detectors \cite{tinto}.
I restrict my attention to the simplest case in which
the source direction is known. 

In order to infer the value of the NP parameters from the 
measurements of the excited vibrational modes of the sphere,
I decompose $E_{ij}$ in terms of spherical harmonics.
In fact, the experimental measurements  give the vibrational 
amplitudes
of the sphere modes which are also naturally expanded in the 
above basis. The use of the same basis makes the connection 
between the NP parameters
and the measured amplitudes straightforward. In formulae
\be
E_{ij}(t)=\sum_{l,m} c_{l,m}(t) S_{ij}^{(l,m)} 
\label{decompo}
\ee
where $S_{ij}^{(0,0)}\equiv \delta_{ij}/\sqrt{4\pi}$
(with $\delta _{ij}$ the Kronecker symbol) and $S_{ij}^{(2,m)}$ 
($m=-2,..2$) 
are five linearly independent symmetric and traceless matrices
such as
\be
S_{ij}^{(l,m)}n^in^j=Y_{lm},\quad l=0,2
\label{nuovaeq}
\ee
The vector
$n_i$ in eqs. (\ref{nuovaeq}) has been defined after 
eq. (\ref{contorno}).

Taking the scalar product I find
\bea
c_{0,0}(t)&=&\frac{4\pi}{3}S_{ij}^{0,0}E_{ij}(t)\nonumber\\
c_{2,m}(t)&=&\frac{8\pi}{15}S_{ij}^{2,m}E_{ij}(t)\label{coccia}
\eea
and then for the NP parameters 
\bea
\Phi_{22} &=& \sqrt{\frac{5}{16\pi}}c_{2,0}(t)
-\sqrt{\frac{1}{4\pi}}c_{0,0}(t)\quad
\Psi_2=-\frac{1}{12}\sqrt{\frac{5}{\pi}}c_{2,0}(t)
-\frac{1}{12}\sqrt{\frac{1}{\pi}}c_{0,0}(t)\cr
Re\Psi_4 &=& -\sqrt{\frac{15}{32\pi}}[c_{2,2}+c_{2,-2}]\quad
Im\Psi_4=-i\sqrt{\frac{15}{32\pi}}[c_{2,2}-c_{2,-2}]\cr
Re\Psi_3 &=& \frac{1}{16}\sqrt{\frac{15}{\pi}}[c_{2,1}-c_{2,-1}]
\quad
Im\Psi_3=\frac{i}{16}\sqrt{\frac{15}{\pi}}[c_{2,1}+c_{2,-1}]
\label{relazionenpc}
\eea
Eqs. (\ref{relazionenpc}) relate
the measurable quantities $c_{l,m}$ with the GW polarization 
states, described
by the NP parameters. 
Eq. (\ref{relazionenpc}) can be put
in correspondence with the output of experimental measurements if 
the $c_{l,m}$ are substituted with their Fourier components at 
the quadrupole and monopole resonant frequencies which, for
the sake of simplicity, I collectively denote by $\omega_0$. 
The $c_{l,m}(\omega_0)$ can be determined in the following way:
once the Fourier amplitudes $A_N(\omega_0)$ are measured, 
by Fourier
transforming (\ref{forzate}) and (\ref{forza}) I get the
Riemann amplitudes $E_{ij}(\omega_0)$ which, using (\ref{coccia}),
yield the desired result.

In order to determine the $A_N(\omega_0)$ amplitudes 
from a given GW signal the following
two conditions must be fulfilled:
\begin{itemize}
\item
the vibrational states of the five-fold degenerate quadrupole 
and monopole modes must be determined. The quadrupole modes 
can be studied by properly combining
the outputs of a set of at least five motion sensors placed
in independent positions on the 
sphere surface.
Explicit formulas for practical and elegant configurations 
of the motion sensors have been reported by
various authors \cite{zhou, merko}.
The vibrational state of the monopole mode
is provided directly by the output of any of the above mentioned
motion sensors.
If resonant motion sensors 
are used, since the quadrupole and monopole states 
resonate at different frequencies, a sixth sensor is needed.
\item
The spectrum of the GW signal must be sufficiently 
broadband to overlap with the antenna quadrupole and monopole 
frequencies. 
\end{itemize}
\section{Gravitational Wave Radiation 
in the Jordan-Brans-Dicke Theory }
\setcounter{equation}{0}
In this section I analyze the signal emitted by a 
compact binary system in the Jordan-Brans-Dicke theory.
I compute the scalar and tensor components of the power radiated 
by the source and study the scalar waveform. 
Eventually I consider the
detectability of the scalar component of the radiation by interferometers and resonant-mass detectors.
\subsection{Scalar and Tensor GWs in the JBD Theory}

In the Jordan-Fierz frame, in which the scalar field 
mixes with the metric but decouples from matter,
the action reads \cite{bd} 
\bea 
S &=& S_{\rm grav}[\phi,g_{\mu\nu}]+S_{\rm m}[\psi_m,g_{\mu\nu}] \nn \\ 
&=& {c^3\over 16\pi}\int d^4x\sqrt{-g}\left[\phi R-
{\omega_{_{BD}}\over\phi}g^{\mu\nu}\partial_\mu\phi\partial_\nu\phi\right]
+{1\over c} \int d^4x L_{\rm m}[\psi_{\rm m},g_{\mu\nu}] \quad ,
\label{uno}
\eea
where $\omega_{_{BD}}$ is a dimensionless constant, whose lower bound
is fixed to be $\omega_{_{BD}}\approx 600$ by experimental data \cite{mssr},
$g_{\mu\nu}$ is the metric tensor, $\phi$ is a scalar field, and $\psi_{\rm m}$
collectively denotes the matter fields of the theory. 

As a preliminary analysis, I perform a weak field approximation around the
background given by a
Minkowskian metric and a constant expectation value for the scalar field 
\bea  
g_{\mu\nu}&=&\eta_{\mu\nu}+h_{\mu\nu} \nn \\
\f &=&\f_0+\xi \quad .
\label{due}
\eea
The standard parametrization $\f_0=2(\o_{_{BD}}+2)/G(2\o_{_{BD}}+3)$, with
$G$ the Newton constant, reproduces GR in the limit
$\omega_{_{BD}}\rightarrow\infty$, which implies $\f_0\rightarrow {1/G}$. 
Defining the new field
\be
\th_{\mu\nu} = h_{\mu\nu} - {1\over 2}\eta_{\mu\nu}h - 
\eta_{\mu\nu}{\xi\over \f_0}
\label{tre}
\ee 
where $h$ is the trace of the fluctuation $h_{\mu\nu}$, 
and choosing the gauge
\be  
\p_\mu \th^{\mu\nu}=0 
\label{quattro}
\ee 
one can write the field equations in the following form
\bea 
\label{cinque} 
\p_\a\p^\a\th_{\mu\nu} &=& -{16\pi\over \f_0}\ \tau_{\mu\nu}\\
\p_\a\p^\a\xi &=&{8\pi\over 2\o_{_{BD}}+3}\ S 
\label{sei} 
\eea 
where 
\bea \label{sette} 
\tau_{\mu\nu} &=& {1\over \f_0} (T_{\mu\nu}+t_{\mu\nu})\\
S &=& - {T\over 2(2\o_{_{BD}}+3)} 
\left(1 - {1\over 2}\th - 2 {\xi\over \f_0}\right) - 
{1\over 16\pi} \left[{1\over 2}\p_\a(\th \p^\a \xi) + 
{2\over \f_0} \p_a (\xi \p^\a \xi)\right] 
\label{otto} 
\eea 
In the equation (\ref{sette}), $T_{\mu\nu}$ is the matter stress-energy 
tensor and $t_{\mu\nu}$ is the gravitational stress-energy pseudo-tensor, 
that is a function of quadratic order in the weak gravitational 
fields $\th_{\mu\nu}$ and $\xi$. 
The reason why I have written the field equations at the quadratic order 
in $\th_{\mu\nu}$ and $\xi$ is that in this way, as I will see later, 
the expressions for $\th_{\mu\nu}$ and $\xi$ include all the terms of order 
$(v/c)^2$, where $v$ is the typical velocity of the source 
(Newtonian approximation). 
 
Let us now compute $\tau^{00}$ and $S$ at the order $(v/c)^2$. 
Introducing the Newtonian potential $U$ produced by the rest-mass 
density $\rho$ 
\be 
U(\vec x, t) = 
\int {\rho(\vec x', t)\over \mid\vec x - \vec x'\mid}\ d^3 x'
\label{nove} 
\ee 
the total pressure $p$ and the specific energy density $\Pi$ (that is the ratio of 
energy density to rest-mass density) I get 
(for a more detailed derivation, see \cite{will}): 
\bea \label{dieci} 
\tau^{00} &=& {1\over \f_0} \rho \quad , \\ 
S &\simeq & - {T\over 2(2\o_{_{BD}}+3)} 
\left(1 - {1\over 2}\th - 2{\xi\over \f_0}\right) \nn \\ 
&=& {\rho\over 2(2\o_{_{BD}}+3}
\left(1+\Pi-3\ {p\over \rho} + {2\o_{_{BD}}+1\over \o_{_{BD}} +2}\ U\right) 
\label{undici} 
\eea 
  
Far from the source, the equations (\ref{cinque}) and (\ref{sei}) 
admit wave--like solutions, which are superpositions of terms of the form 
\bea \label{dodici}
\th_{\mu\nu}(x) &=& A_{\mu\nu}(\vec x, \o) \exp(ik^\a x_\a) + c.c. \\
\xi(x) &=& B(\vec x, \o) \exp(ik^\a x_\a) + c.c. \label{tredici}
\eea 
Without affecting the gauge condition (\ref{quattro}), one can impose
$h=-2\xi/\f_0$ (so that $\th_{\mu\nu} = h_{\mu\nu}$). 
Gauging away the superflous components, one can write the amplitude
$A_{\mu\nu}$ in terms of the three degrees of freedom corrisponding 
to states with helicities $\pm 2$ and 0 \cite{lee}. 
For a wave travelling in the $z$-direction, one thus obtains
\be   
A_{\mu\nu}=\pmatrix{0&0&0&0\cr 0&e_{11}-b&e_{12} &0\cr 0&e_{12}
&-e_{11}-b&0\cr 0&0&0&0\cr},
\label{quattordici}
\ee 
where $b=B/\f_0$.

\subsection{Power emitted in GWs}

The power emitted by a source in GWs depends on the stress-energy 
pseudo-tensor $t^{\mu\nu}$ according to the following expression 
\be \label{quindici}
P_{GW} = r^2 \int \Phi d\O = r^2 \int <t^{0k}>\ \hat x_k\ d\O  
\ee 
where $r$ is the radius of a sphere which contains the source, 
$\O$ is the solid angle, $\Phi$ is the energy flux and the symbol 
$<...>$ implies an 
average over a region of size much larger than the wavelength 
of the GW. At the quadratic order in the weak fields I find
\be   
<t_{0z}> = -\hat z{\f_0 c^4\over 32\pi}
\left[{4(\o_{_{BD}}+1)\over \f_0^2}<(\p_0\xi)(\p_0\xi)> + 
<(\p_0 h_{\a\b})(\p_0 h^{\a\b})>\right]. 
\label{sedici}
\ee
Substituting (\ref{dodici}), (\ref{tredici}) into (\ref{sedici}), one gets 
\be \label{diciassette}  
<t_{0z}> = -\hat z {\f_0 c^4 \o^2 \over 16\pi}
\left[{2(2\o_{_{BD}}+3)\over \f_0^2} \mid B\mid^2 + 
A^{\a\b *}A_{\a\b}-{1\over 2}\mid {A^\a}_\a \mid^2\right],
\ee 
and using (\ref{quattordici})
\be \label{diciotto}  
<t_{0z}> = -\hat z {\f_0 c^4 \o^2\over 8\pi}
\left[\mid e_{11}\mid^2+\mid e_{12}\mid^2 + 
({2\o_{_{BD}}+3})\mid b \mid^2\right].
\ee 
From (\ref{diciotto}) I see that the purely scalar contribution,
associated to $b$, and the traceless tensorial contribution, associated to
$e_{\mu\nu}$, are completely decoupled and can thus be 
treated independently.
\subsection{Power emitted in scalar GWs} 

I now rewrite the scalar wave solution (\ref{tredici}) in the following 
way
\be \label{ventiquattro}
\xi(\vec x, t) = \xi(\vec x, \o) e^{-i\o t} + c. c. 
\ee 
{\it In vacuo}, the spatial part of the previous solution 
(\ref{ventiquattro}) satisfies the Helmholtz equation 
\be \label{venticinque}
(\n^2+\o^2) \xi(\vec x, \o) = 0 
\ee 
The solution of (\ref{venticinque}) can be written as 
\be \label{ventisei} 
\xi(\vec x, \o) = \sum_{jm} X_{jm} h_j^{(1)}(\o r) Y_{jm}(\th, \f) 
\ee 
where $h_j^{(1)}(x)$ are the spherical Hankel functions of the first 
kind, $r$ is the distance of the source from the observer, 
$Y_{jm}(\th, \f)$ are the scalar spherical harmonics and 
the coefficients $X_{jm}$ give the amplitudes of the various 
multipoles which are present in the scalar radiation field.
Solving the inhomogeneous wave equation (\ref{sei}), I find
\be \label{ventisette}   
X_{jm} = 16\pi i \o 
\int_V j_l(\o r') Y_{lm}^*(\th, \f) S(\vec x, \o)\ dV
\ee 
where $j_l(x)$ are the spherical Bessel functions and $r'$ is a radial 
coordinate which assumes its values in the volume $V$ occupied by 
the source. 

Substituting (\ref{sedici}) in (\ref{quindici}), considering the 
expressions (\ref{ventiquattro}) and (\ref{ventisei}), and  
averaging over time, one finally obtains 
\be \label{ventotto}
P_{scal} = {(2\o_{_{BD}}+3) c^4\over 8\pi \f_0} 
\sum_{jm} \mid X_{jm}\mid^2 
\ee 
To compute the power radiated in scalar GWs, one has 
to determine the coefficients $X_{jm}$, defined in (\ref{ventisette}). 
The detailed calculations can be found in the appendix A
of the third reference in \cite{fuc}, while 
here I only give the final results. Introducing the reduced 
mass of the binary system $\mu=m_1 m_2/m$ and the gravitational 
self-energy for the body $a$ (with $a= 1, 2$)
\be \label{ventinove} 
\O_a = - {1\over 2} \int_{V_a} {\rho(\vec x) \rho(\vec x') 
\over \mid \vec x -\vec x'\mid}\ d^3 xd^3x' 
\ee    
one can write the Fourier components with frequency $n\o_0$ in the
Newtonian approximation  
\be\label{trenta}
(X_{00})_n = - {16\sqrt{2\pi}\over 3} {i\o_0 \f_0\over \o_{_{BD}} +2} 
{m \mu \over a}\ n J_n(ne) \\
\ee\newpage
\bea\label{trentabis}
(X_{1\pm 1})_n &=& -\sqrt{2\pi\over 3} 
{2i{\o_0}^2 \f_0 \over \o_{_{BD}} + 2} 
\left({\O_2\over m_2} - {\O_1\over m_1}\right)\ \mu a \nn \\
&& \left[\pm J_n'(ne) - {1\over e} (1-e^2)^{1/2} J_n(ne)\right] \\   
\label{trentuno}
(X_{20})_n &=& {2\over 3} \sqrt{\pi\over 5} 
{i{\o_0}^3 \f_0 \over \o_{_{BD}} + 2} 
\mu a^2 n J_n(ne) \\ 
\label{trentadue}
(X_{2\pm 2})_n &=& \mp 2 \sqrt{\pi\over 30} 
{i{\o_0}^3 \f_0 \over \o_{_{BD}} + 2} \mu a^2 \nn \\ 
&& {1\over n} 
\{(e^2-2) J_n(ne)/(ne^2) + 2 (1-e^2) J_n'(ne)/e \nn \\ 
&\mp& 2(1-e^2)^{1/2} [(1-e^2) J_n(ne)/e^2 - J_n'(ne)/(ne)]\} 
\eea    
Substituting these expressions in (\ref{ventotto}), leads to the power 
radiated in scalar 
GWs in the $n$-th harmonic 
\be \label{trentatre}
(P_{scal})_n = P_n^{j=0} + P_n^{j=1} + P_n^{j=2} 
\ee 
where the monopole, dipole and quadrupole terms are respectively 
\bea \label{trentaquattro} 
P_n^{j=0} &=& {64\over 9(\o_{_{BD}} + 2)} 
{m^3 \mu^2 G^4 \over a^5 c^5} n^2 J_n^2(ne) \nn \\
&=& {64\over 9(\o_{_{BD}} + 2)} 
{m^3 \mu^2 G^4 \over a^5 c^5}\ m(n; e) \\
\label{trentacinque}
P_n^{j=1} &=& {4\over 3(\o_{_{BD}} +2)} 
{m^2 \mu^2 G^3 \over a^4 c^3} 
\left({\O_2\over m_2} - {\O_1\over m_1}\right)^2 \nn \\ 
&& n^2 \left[ J_n'^2(ne) + {1\over e^2} (1-e^2) J_n^2(ne)\right] \nn \\ 
&=& {4\over 3(\o_{_{BD}} +2)} 
{m^2 \mu^2 G^3 \over a^4 c^3} 
\left({\O_2\over m_2} - {\O_1\over m_1}\right)^2 d(n; e) \\ 
\label{trentasei} 
P_n^{j=2} &=&  {8\over 15(\o_{_{BD}} +2)} 
{m^3\mu^2 G^4\over a^5 c^5}\ g(n; e) 
\eea 

The total power radiated in scalar GWs by a binary system is 
the sum of three terms 
\be \label{trentaseibis}
P_{scal} = P^{j=0} + P^{j=1} + P^{j=2} 
\ee 
where 
\bea \label{trentasette} 
P^{j=0} &=& {16\over 9(\o_{_{BD}}+2)} 
{G^4\over c^5} {m_1^2 m_2^2 m\over a^5} 
{e^2\over (1-e^2)^{7/2}} \left(1+{e^2\over 4}\right) \\ 
\label{trentotto} 
P^{j=1} &=& {2\over \o_{_{BD}}+2}
\left({\O_2\over m_2} - {\O_1\over m_1}\right)^2  
{G^3\over c^3} {m_1^2 m_2^2 \over a^4} 
{1\over (1-e^2)^{5/2}} \left(1+{e^2\over 2}\right) \\
\label{trentanove}  
P^{j=2} &=& {8\over 15(\o_{_{BD}}+2)} 
{G^4\over c^5} {m_1^2 m_2^2 m\over a^5} 
{1\over (1-e^2)^{7/2}} \left(1+{73\over 24} e^2 + {37\over 96} e^4\right) 
\eea 
Note that $P^{j=0}, P^{j=1}, P^{j=2}$ all  go to zero in the limit $\o_{_{BD}} 
\to \infty$.

\subsection{Scalar GWs} 

I now give the explicit form of the scalar GWs
radiated by a binary system. To this end, note that 
the major semi-axis, $a$, is related to the total energy, 
$E$, of the system through the following equation
\be \label{quaranta} 
a = - {Gm_1 m_2 \over 2E} 
\ee 
Let us consider the case of a circular orbit, remembering that in the 
last phase of evolution of a binary system this condition is usually 
satisfied. Furthermore I will also assume $m_1 = m_2$.
With these positions
only the quadrupole term, (\ref{trentasei}), of the gravitational 
radiation is different from zero. 
The total power radiated in GWs, averaged over time, 
is then given by (\ref{trentasette})-(\ref{trentanove})
\be \label{quarantuno} 
P = {8\over 15(\o_{_{BD}}+2)} {G^4\over c^5} {m_1^2 m_2^2 m\over d^5} 
[6(2\o_{_{BD}}+3) + 1] 
\ee
where $d$ is the relative distance between the two stars. 
The time variation of $d$ in one orbital period is 
\be \label{quarantadue} 
\dot d = - {G m_1 m_2\over 2 E^2} P
\ee 
Finally, substituting (\ref{quaranta}), (\ref{quarantuno})
in (\ref{quarantadue}) and integrating over time, one obtains 
\be \label{quarantatre}
d = 2 \left({2\over 15}  
{12\o_{_{BD}}+19\over \o_{_{BD}}+2}
{G^3 m_1 m_2 m\over c^5} \right)^{1/4} \tau^4
\ee 
where I have defined $\tau= t_c - t$, $t_c$ being the time of
the collapse between the two bodies. 

From (\ref{ventisei}), (\ref{trenta})-(\ref{trentadue}) and 
one can deduce the form of the scalar field (see appendix B 
of the third in \cite{fuc} for 
details) which, for equal masses, is 
\be \label{quarantaquattro} 
\xi(t) = - {2 \mu\over r(2\o_{_{BD}}+3)} 
\left[v^2 + {m\over d} - (\hat n\cdot \vec v)^2 + 
{m\over d^3} (\hat n\cdot \vec d)\right] 
\ee 
where $r$ is the distance of the source from the observer, 
and $\hat n$ is the versor of the line of sight from the observer to the 
binary system center of mass. Indicating with $\g$ the inclination angle,
that is the angle between the orbital plane and the reference plane
(defined to be a plane perpendicular to the line of sight), and
with $\psi$ the true anomaly, that is the angle between $d$ and the $x$-axis
in the orbital plane $x$-$y$, yields $\hat n \cdot \vec d = d \sin\g\sin\psi$. 
Then from (\ref{quarantaquattro}) one obtains 
\be \label{quarantacinque} 
\xi(t) = {2G\mu m\over (2\o_{_{BD}}+3)c^4 d r} \sin^2\g\cos(2\psi(t))
\ee 
which can also be written as
\be \label{quarantasei} 
\xi(\tau) = \xi_0(\tau) \sin(\chi(\tau)+\bar\chi)
\ee
where $\bar\chi$ is an arbitrary phase and the amplitude $\xi_0(\tau)$ 
is given by 
\bea \label{quarantasette} 
\xi_0(\tau) &=& {2G\mu m\over (2\o_{_{BD}}+3)c^4 d r} \sin^2\g \nn \\
&=& {1\over 2(2\o_{_{BD}}+3) r} 
\left({\o_{_{BD}}+2\over 12\o_{_{BD}}+19}\right)^{1/4} 
\left({15G\over 2c^{11}}\right)^{1/4}
{{M_c}^{5/4}\over \tau^{1/4}} \sin^2\g
\eea 
In the last expression, I have introduced the definition of 
the chirp mass $M_c = (m_1m_2)^{3/5}/m^{1/5}$.   

\subsection{Detectability of the scalar GWs}

Let me now study the interaction of the scalar GWs a spherical GW 
detector.

As usual, I characterize the sensitivity of the detector by the spectral 
density of strain $S_h(f)$ $[\hbox{Hz}]^{-1}$. The optimum performance of a 
detector is obtained by 
filtering  the output with a filter matched to the signal. The energy 
signal-to-noise ratio $SNR$ 
of the filter output is given by the well-known formula:
\begin{equation}
SNR = \int^{+\infty}_{-\infty}\frac{|H(f)|^2} {S_{h}(f)}\,df
\label{snr}
\end{equation}
where $H(f)$ is the Fourier transform of the scalar 
gravitational  waveform $h_s(t)=G\/\xi_0(t)$.

I must now take into account the astrophysical
restrictions on the validity of the waveform (\ref{quarantasei}) 
which is obtained in the Newtonian approximation for point-like
masses.
In the following, I will take the point of view that this approximation
breaks down when there are five cycles remaining to collapse \cite{dewey,cf}.

The five-cycles limit will be used to restrict the range of $M_c$
over which my analysis will be performed. From (\ref{quarantatre}),
one can obtain
\bea \label{quarantotto}
\o_g(\tau) &=& 2\o_0 = 2 \sqrt{Gm\over d^3} \nn \\
&=& 2\left({15c^5\over 64G^{5/3}}\right)^{3/8}
\left({\o_{_{BD}} + 2\over 12\o_{_{BD}} + 19}\right)^{3/8}
{1\over {M_c}^{5/8}} \tau^{3/8}
\eea
Integrating (\ref{quarantotto}) yields the amount of phase
until coalescence
\be \label{quarantanove}
\chi(\tau) = {16\over 5}\left({15c^5\over 64G^{5/3}}\right)^{3/8}
\left({\o_{_{BD}} + 2\over 12\o_{_{BD}} + 19}\right)^{3/8}
\left({\tau \over {M_c}}\right)^{5/8}
\ee
Setting (\ref{quarantanove}) equal to the limit period,
$T_{5~cycles}=5(2\pi)$, solving for $\tau$
and using (\ref{quarantotto}) leads to
\be \label{cinquanta}
\o_{5~cycles} = 2\pi (6870\ \hbox{Hz})
\left({\o_{_{BD}} + 2\over 12\o_{_{BD}} + 19}\right)^{3/5}
{M_\odot\over M_c}
\ee
Taking $\o_{_{BD}} = 600$, the previous limit reads
\be \label{cinquantuno}
\o_{5~cycles} = 2\pi (1547\ \hbox{Hz})
{M_\odot\over M_c}
\ee
A GW excites those vibrational modes of a resonant body having the proper simmetry. 
In the framework of the JBD theory the spheroidal modes with 
$l=2$ and $l=0$ are sensitive to the incoming GW. Thanks to its
multimode nature, a single sphere is capable of detecting GW's from all 
directions and polarizations.
I now evaluate the $SNR$ of a resonant-mass detector of spherical shape for its 
quadrupole mode with $m=0$ and its monopole mode.
In a resonant-mass detector, $S_h(f)$ is a resonant curve and can
be characterized by its value at resonance $S_{h}(f_n)$
and by its half height width \cite{pizz}. $S_{h}(f_n)$ can thus
be written as
\begin{equation}
  S_h(f_n) = \frac{G}{c^3}\frac{4kT}{\sigma_{n} Q_{n} f_{n}}
\label{esseacca}
\end{equation}

Here $\sigma_{n}$ is the cross-section associated with the $n\/$-th
resonant mode,
$T$ is the thermodynamic temperature of the detector 
and $Q_n\/$ is the quality factor of the mode.

The half height width of $S_{h}(f)$ gives the bandwidth of the resonant
mode

\begin{equation}
\Delta f_n = \frac{f_n}{Q_n} \Gamma_{n} ^{-1/2}
\label{deltaeffe}
\end{equation}

Here, $\Gamma_n$ is the ratio of the wideband noise in the $n\/$-th resonance
bandwidth to the narrowband noise. 

From the resonant-mass detector viewpoint, the chirp signal can be treated as a 
transient GW, depositing energy in a time-scale short with respect to the 
detector damping time. I can then consider constant the Fourier transform of 
the waveform within the band of the detector and write \cite{pizz}
\be \label{snr0}
SNR = \frac{2\pi\Delta f_{n} |H(f_n)|^{2}}{S_{h}(f_n)}
\ee

The cross-sections associated to the vibrational modes
with $l=0$ and $l=2$, $m=0$ are respectively \cite{fuc}
\bea \label{cinquantaquattro}
\s_{(n0)} &=& H_n {G M {v_s}^2\over c^3 (\o_{_{BD}}+2)}
\\ 
\label{cinquantacinque}
\s_{(n2)} &=& {F_n\over 6} {G M {v_s}^2\over c^3 (\o_{_{BD}}+2)}
\eea
All parameters entering the previous equation refer to the detector
$M$ is its mass, $v_s$ the sound velocity
and the constants $H_n$ and $F_n$ are given in \cite{fuc}.
The signal-to noise ratio can be
calculated analytically by approximating the waveform with
a truncated Taylor expansion around $t=0$,
where $\o_g(t=0) = \o_{nl}$ \cite{clark,dewey}
\be \label{cinquantasette}
h_s(t) \approx G \xi_0(t=0)
\sin\left[\o_{nl} t + {1\over 2} \left({d\o\over dt}\right)_{t=0} t^2\right]
\ee
Using quantum limited readout systems, one finally obtains
\bea \label{cinquantanove}
(SNR_n)_{l=0} &=&
{5\cdot 2^{1/3} H_n G^{5/3}  \over
32(\o_{_{BD}} +2)(12\o_{_{BD}} + 19) \hbar c^3} \nn \\ 
&& {{M_c}^{5/3} M {v_s}^2 \over r^2 {\o_{n0}}^{4/3}}
\sin^4\g \\
\label{sessanta}
(SNR_n)_{l=2} &=&
{5\cdot 2^{1/3} F_n G^{5/3}  \over
192(\o_{_{BD}} +2)(12\o_{_{BD}} + 19) \hbar c^3} \nn \\ 
&& {{M_c}^{5/3} M {v_s}^2 \over r^2 {\o_{n0}}^{4/3}}
\sin^4\g
\eea
which are respectively the signal-to-noise ratio for the modes
with $l=0$ and $l=2$, $m=0$ of a spherical detector.

It has been proposed to realize spherical detectors with 3 meters diameter,
made of copper alloys, with mass of the order of 100 tons \cite{fc}.
This proposed detector has resonant frequencies of  $\omega_{12} = 2 \pi \cdot
807$
rad/s and $\omega_{10} = 2 \pi \cdot 1655$ rad/s.
In the case of optimally oriented orbits (inclination angle $\g = \pi/2$) and 
$\o_{_{BD}}=600$, the inspiralling of two compact objects of 1.4 solar masses 
each will then be detected with $SNR =1$ up to a source distance 
$r(\o_{10})\simeq 30$ kpc and $r(\o_{12})\simeq 30$ kpc.

\section{The hollow sphere}
An appealing variant of the massive sphere is a {\it hollow\/} sphere
\cite{vega-98}. The latter has the remarkable property that it enables the
detector to monitor GW signals in a significantly {\it lower frequency range\/}
---down to about 200 Hz--- than its massive counterpart for comparable
sphere masses. This can be considered a positive advantage for a future
world wide network of GW detectors, as the sensitivity range of such
antenna overlaps with that of the large scale interferometers, now in
a rather advanced state of construction \cite{ligo,virgo}. In this Section I study the response of such a detector to the GW energy emitted by a binary system constituted of stars of masses of the order of the solar mass. 
A hollow sphere obviously has the same symmetry of the massive one, so
the general structure of its {\it normal modes\/} of vibration 
is very similar\cite{vega-98} to that of the solid sphere. 
In particular, the hollow sphere is
very well adapted to sense and monitor the presence of scalar modes in
the incoming GW signal. The extension
of the analysis of the previous Sections to a hollow sphere
is quite straightforward and in the following I will only give the main results. Due to the different geometry, the vibrational modes of a hollow sphere differ from those studied in Section 2. 
In the case of a hollow
sphere, I have two boundaries given by
the outer and the inner surfaces
of the solid itself. I use the notation $a\/$ for the inner radius, and $R\/$ for
the outer radius. The boundary conditions are thus expressed by
\begin{equation}
\sigma_{ij}n_j=0\hspace{1cm}\mbox{at}\hspace{0.4cm} r=R
\hspace{0.4cm}\mbox{and at}\hspace{0.4cm}r=a\hspace{0.5cm}(R\geq a \geq 0),
\label{bc}
\end{equation}
(\ref{eqmotomio}) must now be solved subject to this 
boundary conditions. The solution that leads to spheroidal modes
is still (\ref{quattrostagioni}) where the radial functions $A_{nl}(r)$ and $B_{nl}(r)$ have rather
complicated expressions:

\begin{eqnarray}
\label{1.8}
A_{nl}(r) &=& C_{nl}\,\left[\frac{1}{q_{nl}^S}\,\frac{d}{dr}\,
j_l(q_{nl}^Sr) -
l(l+1)\,K_{nl}\,\frac{j_l(k_{nl}^Sr)}{k_{nl}^Sr}+\right.
  \nonumber \\
&& \ \qquad\quad + \left. D_{nl}\,\frac{1}{q_{nl}^S}\,\frac{d}{dr}\,
y_l(q_{nl}^Sr)
- l(l+1)\,\tilde D_{nl}\,\frac{y_l(k_{nl}^Sr)}{k_{nl}^Sr}\right]
\label{1.8a}  \\
B_{nl}(r) &=& C_{nl}\,\left[\frac{j_l(q_{nl}^Sr)}{q_{nl}^Sr} -
K_{nl}\,\frac 1{k_{nl}^Sr}\,\frac d{dr}\left\{r\,j_l(k_{nl}^Sr)\right\} +
\right.  \nonumber \\
&& \ \qquad\quad + \left. D_{nl}\,\frac{y_l(q_{nl}^Sr)}{q_{nl}^Sr} -
\tilde D_{nl}\,\frac 1{k_{nl}^Sr}\,\frac d{dr}\left\{r\,y_l(k_{nl}^Sr)\right\}
\right] \label{1.8b}
\end{eqnarray}

Here $k_{nl}^SR$ and $q_{nl}^SR$ are dimensionless {\it eigenvalues\/},
and they are the solution to a rather complicated algebraic equation for
the frequencies $\omega\/$\,=\,$\omega_{nl}\/$ ---see
\cite{vega-98} for details. In (\ref{1.8a}) and (\ref{1.8b}) I have set

\begin{equation}
K_{nl}\equiv\frac{C_{\rm t}q_{nl}^S}{C_{\rm l}k_{nl}^S}\ ,\qquad
D_{nl}\equiv\frac{q_{nl}^S}{k_{nl}^S}\,E\ ,\qquad
\tilde D_{nl}\equiv\frac{C_{\rm t}Fq_{nl}^S}{C_{\rm l}k_{nl}^S}
\label{1.9}
\end{equation}
and introduced the normalisation constant $C_{nl}$, which is fixed by the
orthogonality properties

\begin{equation}
\int_V({\bf u}_{n^{\prime}l^{\prime}m^{\prime}}^S)^*\cdot
({\bf u}_{nlm}^S)\,\varrho_0\,d^3 x =
M\,\delta_{nn^{\prime}}\delta_{ll^{\prime}}\delta_{mm^{\prime}} 
\label{1.10}
\end{equation}
where $M\/$ is the mass of the hollow sphere:

\begin{equation}
M = \frac{4\pi}3\,\varrho_0 R^3\,(1-\varsigma ^3)\ ,\qquad
\varsigma\equiv\frac{a}{R}\leq 1
\label{1.11}
\end{equation}

Equation (\ref{1.10}) fixes the value of $C_{nl}$ through the radial integral

\begin{equation}
\int_{\varsigma R}^R\,\left[A_{nl}^2(r) + l(l+1)\,B_{nl}^2(r)\right]\,
r^2dr = \frac{4\pi}3\varrho_0\,(1-\varsigma^3)R^3
\label{1.12}
\end{equation}
as can be easily verified by using
well known properties of angular momentum operators and spherical harmonics.
I shall later specify the values of the different parameters appearing in
the above expressions as required in each particular case which will in due
course be considered.
As seen in reference \cite{lobo}, a scalar--tensor theory of GWs such as
JBD predicts the excitation of the sphere's monopole modes {\it as well as
the\/} $m\/$\,=\,0 quadrupole modes. In order to calculate the energy absorbed
by the detector according to that theory it is necessary to calculate the
energy deposited by the wave in those modes, and this in turn requires that
I solve the elasticity equation with the GW driving term included in its
right hand side. The result of such calculation was presented in full
generality in reference \cite{lobo}, and is directly applicable here
because the structure of the oscillation eigenmodes of a hollow sphere is
equal to that of the massive sphere ---only the explicit form of the
wavefunctions needs to be changed. I thus have

\begin{equation}
E_{\rm osc}(\omega_{nl}) = \frac{1}{2}\,Mb^2_{nl}
  \,\sum_{m=-l}^l\,|G^{(lm)}(\omega_{nl})|^2
\label{2.0}
\end{equation}
where $G^{(lm)}(\omega_{nl})$ is the Fourier amplitude of the corresponding
incoming GW mode, and

\begin{eqnarray}
\label{2.1}
b_{n0} &=& -\frac{\varrho_0}{M}\,\int_a^R\,A_{n0}(r)\,r^3 dr
\label{2.1a} \\[1 ex]
b_{n2} &=& -\frac{\varrho_0}{M}\,\int_a^R\,\left[A_{n2}(r)
	 + 3B_{n2}(r)\right]\,r^3 dr
\label{2.1b}
\end{eqnarray}
for monopole and quadrupole modes, respectively, and $A_{nl}(r)$ and
$B_{nl}(r)$ are given by (\ref{1.8}). Explicit calculation yields

\begin{eqnarray}
\label{2.2}
\frac{b_{n0}}{R} &=& \frac 3{4\pi}\,\frac{C_{n0}}{1-\varsigma^3}\,
\left[\Lambda(R) - \varsigma^3\Lambda(a)\right] \label{2.2a} \\[1 ex]
\frac{b_{n2}}{R} &=& \frac 3{4\pi}\,\frac{C_{n2}}{1-\varsigma^3}\,
\left[\Sigma(R) - \varsigma^3\Sigma(a)\right] \label{2.2b}
\end{eqnarray}
with

\begin{eqnarray}
\label{2.3}
\Lambda(z) & \equiv & \frac{j_2(q_{n0}z)}{q_{n0}R} +
 D_{n0}\,\frac{y_2(q_{n0}z)}{q_{n0}R}   \label{2.3a} \\[1 em]
\Sigma(z) & \equiv & \frac{j_2(q_{n2}z)}{q_{n2}R} -
3K_{n2}\,\frac{j_2(k_{n2}z)}{k_{n2}R} +
D_{n2}\,\frac{y_2(q_{n2}z)}{q_{n2}R} -
3\tilde D_{n2}\,\frac{y_2(k_{n2}z)}{k_{n2}R}   \label{2.3b}
\end{eqnarray}

The absorption {\it cross section\/}, defined as the ratio of the absorbed
energy to the incoming flux, can be calculated thanks to an {\it optical
theorem\/}, as proved e.g.\ by Weinberg \cite{wein-72}. According to that
theorem, the absorption cross section for a signal of frequency $\omega\/$
close to $\omega_N\/$, say, the frequency of the detector mode excited by
the incoming GW, is given by the expression

\begin{equation}
\sigma(\omega) = \frac{10\,\pi\eta c^2}{\omega^2}\,
		 \frac{\Gamma^2/4}{(\omega -\omega_N)^2 + \Gamma^2/4}
\label{2.4}
\end{equation}
where $\Gamma$ is the {\it linewitdh\/} of the mode ---which can be
arbitrarily small, as assumed in the previous section---, and $\eta\/$
is the dimensionless ratio

\begin{equation}
  \eta = \frac{\Gamma_{\rm grav}}{\Gamma} =
	 \frac{1}{\Gamma}\,\frac{P_{GW}}{E_{\rm osc}}
\label{2.5}
\end{equation}
where $P_{GW}$ is the energy {\it re-emitted\/} by the detector in the form
of GWs as a consequence of its being set to oscillate by the incoming signal. In the following I will only consider the case 
$P_{GW}=P_{\rm scalar-tensor}$ with \cite{lobo,fuc}
\begin{equation}
P_{\rm scalar-tensor} = \frac{2G\,\omega ^6}{5c^5\,(2\omega _{BD}+3)}\,
\left[\left|Q_{kk}(\omega)\right|^2 +
\frac 13\,Q_{ij}^*(\omega)Q_{ij}(\omega)\right]
\label{2.8}
\end{equation}
where $Q_{ij}(\omega)$ is the quadrupole moment of the hollow sphere:

\begin{equation}
Q_{ij}(\omega) = \int_{\rm Antenna}\,x_ix_j\,\varrho({\bf x},\omega)\,d^3x
\end{equation}
and $\omega_{BD}\/$ is Brans--Dicke's parameter.

\section{Scalar-tensor cross sections}

Explicit calculation shows that $P_{\rm scalar-tensor}$ is made up of two
contributions:

\begin{equation}
P_{\rm scalar-tensor} = P_{00} + P_{20}
\label{3.1}
\end{equation}
where $P_{00}$ is the scalar, or monopole contribution to the emitted power,
while $P_{20}$ comes from the central quadrupole mode which, as discussed in
\cite{fuc} and \cite{lobo}, is excited together with monopole in JBD
theory. One must however recall that monopole and quadrupole modes of the
sphere happen at {\it different frequencies\/}, so that cross sections for
them only make sense if defined separately. More precisely,

\begin{eqnarray}
\label{3.2}
\sigma_{n0}(\omega) & = & \frac{10\pi\,\eta_{n0}\,c^2}{\omega^2}\,
  \frac{\Gamma_{n0}^2/4}{(\omega - \omega_{n0})^2 + \Gamma_{n0}^2/4}
  \label{3.2a} \\
\sigma_{n2}(\omega) & = & \frac{10\pi\,\eta_{n2}\,c^2}{\omega^2}\,
  \frac{\Gamma_{n2}^2/4}{(\omega - \omega_{n2})^2 + \Gamma_{n2}^2/4}
  \label{3.2b}
\end{eqnarray}
where $\eta_{n0}$ and $\eta_{n2}$ are defined like in (\ref{2.5}), with all
terms referring to the corresponding modes. After some algebra one finds that

\begin{eqnarray}
\label{3.3}
\sigma_{n0}(\omega) & = & H_n\,\frac{GMv_S^2}{(\omega_{BD}+2)\,c^3}\,
  \frac{\Gamma_{n0}^2/4}{(\omega - \omega_{n0})^2 + \Gamma_{n0}^2/4}
  \label{3.3a} \\
\sigma_{n2}(\omega) & = & F_n\,\frac{GMv_S^2}{(\omega_{BD}+2)\,c^3}\,
  \frac{\Gamma_{n2}^2/4}{(\omega - \omega_{n2})^2 + \Gamma_{n2}^2/4}
  \label{3.3b}
\end{eqnarray}

Here, I have defined the dimensionless quantities

\begin{eqnarray}
\label{3.4}
  H_n & = & \frac{4\pi^2}{ 9\,(1+\sigma_P)}\,(k_{n0}b_{n0})^2	\label{3.4a} \\
  F_n & = & \frac{8\pi^2}{15\,(1+\sigma_P)}\,(k_{n2}b_{n2})^2	\label{3.4b}
\end{eqnarray}
where $\sigma_P\/$ represents the sphere material's Poisson ratio (most
often very close to a value of 1/3), and the $b_{nl}\/$ are defined in
(\ref{2.2}); $v_S\/$ is the speed of sound in the material of the sphere.

In tables \ref{t.1} and \ref{t.2} I give a few numerical values of the
above cross section coefficients.
\begin{table}[ht]
\begin{center}
\footnotesize\rm
\caption{Eigenvalues $k_{n0}^SR$, relative weights $D_{n0}$ and $H_n$
coefficients for a hollow sphere with Poisson ratio $\sigma_P$=1/3.
Values are given for a few different thickness parameters $\varsigma$.}
\label{t.1}
\begin{tabular}{lllll} 
\topline
$\varsigma$ & $n$ &$k_{n0}^SR$ & $D_{n0}$ &$H_n$ \\ 
\midline
0.01 & 1 & 5.48738 & -.000143328 & 0.90929 \\
     & 1 & 12.2332 & -.001.59636 & 0.14194 \\
     & 2 & 18.6321 & -.00558961  & 0.05926 \\
     & 4 & 24.9693 & -.001279    & 0.03267 \\
0.10 & 1 & 5.45410 & -0.014218 & 0.89530 \\
     & 1 & 11.9241 & -0.151377 & 0.15048 \\
     & 2 & 17.7277 & -0.479543 & 0.04922 \\
     & 4 & 23.5416 & -0.859885 & 0.04311 \\
0.15 & 1 & 5.37709 & -0.045574 & 0.86076 \\
     & 2 & 11.3879 & -0.434591 & 0.17646 \\
     & 3 & 17.105  & -0.939629 & 0.05674 \\
     & 4 & 23.605  & -0.806574 & 0.05396 \\
0.25 & 1 & 5.04842 & -0.179999 & 0.73727 \\
     & 2 & 10.6515 & -0.960417 & 0.30532 \\
     & 3 & 17.8193 & -0.425087 & 0.04275 \\
     & 4 & 25.8063 &  0.440100 & 0.06347 \\
0.50 & 1 & 3.96914 & -0.631169 & 0.49429 \\
     & 2 & 13.2369 &  0.531684 & 0.58140 \\
     & 3 & 25.4531 &  0.245321 & 0.01728 \\
     & 4 & 37.9129 &  0.161117 & 0.07192 \\
0.75 & 1 & 3.26524 & -0.901244 & 0.43070 \\
     & 2 & 25.3468 &  0.188845 & 0.66284 \\
     & 3 & 50.3718 &  0.093173 & 0.00341 \\
     & 4 & 75.469  &  0.061981 & 0.07480 \\
0.90 & 1 & 2.98141 & -0.963552 & 0.42043 \\
     & 2 & 62.9027 &  0.067342 & 0.67689 \\
     & 3 & 125.699 &  0.033573 & 0.00047 \\
     & 4 & 188.519 &  0.022334 & 0.07538 \\
\bottomline
\end{tabular}
\end{center}
\end{table}
\begin{table}[ht]
\begin{center}
\footnotesize\rm
\caption{Eigenvalues $k_{n2}^SR\/$, relative weights $K_{n2}$, $D_{n2}$,
$\tilde D_{n2}$ and $F_n\/$ coefficients for a hollow sphere with Poisson
ratio $\sigma_P$\,=\,1/3. Values are given for a few different thickness
parameters $\varsigma$.}
\label{t.2}
\begin{tabular}{lllllll}
\topline

$\ \ \varsigma$ & $n$ & $\ \ k_{n2}^SR$ & \qquad\ \ $K_{n2}$ &
 \qquad\ \ $D_{n2}$ &  \qquad\ \ $\tilde D_{n2}$ & \qquad $F_n$ \\
\midline
0.10 & 1 &  2.63836 &  0.855799 & 0.000395 & -0.003142 & 2.94602 \\
     & 2 &  5.07358 &  0.751837 & 0.002351 & -0.018451 & 1.16934 \\
     & 3 & 10.96090 &  0.476073 & 0.009821 & -0.071685 & 0.02207 \\[1 ex]
0.15 & 1 &  2.61161 &  0.796019 &  0.001174 & -0.009288 & 2.86913 \\
     & 2 &  5.02815 &  0.723984 &  0.007028 & -0.053849 & 1.24153 \\
     & 3 &  8.25809 & -2.010150 & -0.094986 &  0.672786 & 0.08113 \\[1 ex]
0.25 & 1 &  2.49122 &  0.606536 &  0.003210 & -0.02494 & 2.55218 \\
     & 2 &  4.91223 &  0.647204 &  0.019483 & -0.13867 & 1.55022 \\
     & 3 &  8.24282 & -1.984426 & -0.126671 &  0.67506 & 0.05325 \\
     & 4 & 10.97725 &  0.432548 & -0.012194 &  0.02236 & 0.03503 \\[1 ex]
0.50 & 1 &  1.94340 &  0.300212 &  0.003041 & -0.02268 & 1.61978 \\
     & 2 &  5.06453 &  0.745258 &  0.005133 & -0.02889 & 2.29572 \\
     & 3 & 10.11189 &  1.795862 & -1.697480 &  2.98276 & 0.19707 \\
     & 4 & 15.91970 & -1.632550 & -1.965780 & -0.30953 & 0.17108 \\[1 ex]
0.75 & 1 &  1.44965 &  0.225040 &  0.001376 & -0.01017 & 1.15291 \\
     & 2 &  5.21599 &  0.910998 & -0.197532 &  0.40944 & 1.82276 \\
     & 3 & 13.93290 &  0.243382 &  0.748219 & -3.20130 & 1.08952 \\
     & 4 & 23.76319 &  0.550278 & -0.230203 & -0.81767 & 0.08114 \\[1 ex]
0.90 & 1 &  1.26565 &  0.213082 &  0.001019 & -0.00755 & 1.03864 \\
     & 2 &  4.97703 &  0.939420 & -0.323067 &  0.52279 & 1.54106 \\
     & 3 & 31.86429 &  6.012680 & -0.259533 &  4.05274 & 1.46486 \\
     & 4 & 61.29948 &  0.205362 & -0.673148 & -1.04369 & 0.13470 \\
\bottomline
\end{tabular}
\end{center}
\end{table}

As already stressed in reference \cite{vega-98}, one of the main advantages
of a hollow sphere is that it enables to reach good sensitivities at lower
frequencies than a solid sphere. For example, a hollow sphere of the same
material and mass as a solid one ($\varsigma$\,=\,0) has eigenfrequencies
which are smaller by

\begin{equation}
 \omega_{nl}(\varsigma) = \omega_{nl}(\varsigma=0)\,(1-\varsigma^3)^{1/3}
 \label{3.5}
\end{equation}
for any mode indices $n\/$ and $l\/$. I now consider the detectability of
JBD GW waves coming from several interesting sources with a hollow sphere.

The values of the coefficients $F_n\/$ and $H_n\/$, together with the
expressions (\ref{3.2}) for the cross sections of the hollow sphere, can
be used to estimate the maximum distances at which a coalescing compact
binary system and a gravitational collapse event can be seen with such
detector. I consider these in turn.

By taking as a source of GWs a binary system formed by two neutron stars,
each of them with a mass of $m_1$\,=\,$m_2$\,=\,1.4\,$M_\odot$.
The {\it chirp mass\/} corresponding to this system is
$M_c\/$\,$\equiv$\,$(m_1m_2)^{3/5}\,(m_1+m_2)^{-1/5}$\,=\,1.22\,$M_\odot$,
and $\nu_{\rm [5\ cycles]}$\,=\,1270 Hz.
Repeating the analysis
carried on in  Section three I find a formula for the minimum distance at 
which a measurement can be performed given a certain signal to noise ratio (SNR), for a {\it quantum limited\/} detector 

\begin{eqnarray}
\label{4.1}
 r(\omega_{n0}) & = &\left[\frac{5\cdot 2^{1/3}}{32}\,
  \frac{1}{(\Omega_{BD}+2)(12\Omega_{BD}+19)}\,
  \frac{G^{5/3}M_c^{5/3}}{c^3}\,\frac{Mv_S^2}{\hbar\omega_{n0}^{4/3}SNR}\,H_n
  \right]^{1/2}  \label{4.1a}  \\
 r(\omega_{n2}) & = &\left[\frac{5\cdot 2^{1/3}}{192}
  \frac{1}{(\Omega_{BD}+2)(12\Omega_{BD}+19)}\,
  \frac{G^{5/3}M_c^{5/3}}{c^3}\,\frac{Mv_S^2}{\hbar\omega_{n2}^{4/3}SNR}\,F_n
  \right]^{1/2}  \label{4.1b}
\end{eqnarray}

For a CuAL sphere, the speed of sound is $v_S\/$\,=\,4700 m/sec. I report
in table \ref{t.3} the maximum distances at which a JBD binary can be seen
with a 100 ton hollow spherical detector, including the size of the sphere
(diameter and thickness factor) for $SNR=1$. The Brans-Dicke parameter $\Omega_{BD}\/$
has been given a value of 600. This high value has as a consequence that
only relatively nearby binaries can be scrutinised by means of their scalar
radiation of GWs. A slight improvement in sensitivity is appreciated as the
diameter increases in a fixed mass detector. Vacancies in the tables mean
the corresponding frequencies are higher than $\nu_{\rm [5\ cycles]}$.

\begin{table}[ht]
\begin{center}
\footnotesize\rm
\caption{Eigenfrequencies, sizes and distances at which coalescing binaries
can be seen by monitoring of their emitted JBD GWs. Figures correspond to a
100 ton CuAl hollow sphere.}
\label{t.3}
\begin{tabular}{llllll}
\topline
$\varsigma $ & $\Phi$ (m) & $\nu_{10}$(Hz) & $\nu_{12}$ (Hz)
& $r(\nu_{10})$ (kpc) & $r(\nu_{12})$ (kpc) \\ \midline
0.00 & 2.94 & 1655 & 807 & $-$ & 29.8 \\
0.25 & 2.96 & 1562 & 771 & $-$ & 30.3 \\
0.50 & 3.08 & 1180 & 578 & 55  & 31.1 \\
0.75 & 3.5  & 845  & 375 & 64  & 33   \\
0.90 & 4.5  & 600  & 254 & 80  & 40   \\
\bottomline
\end{tabular}
\end{center}
\end{table}

\begin{table}
\begin{center}
\footnotesize\rm
\caption{Eigenfrequencies, sizes and distances at which coalescing binaries
can be seen by monitoring of their emitted JBD GWs. Figures correspond to a
3 metres external diameter CuAl hollow sphere.}
\label{t.4}
\begin{tabular}{llllll}
\topline
$\varsigma $ & $M$ (ton) & $\nu _{10}$(Hz) & $\nu_{12}$(Hz) &
$r(\nu_{10})$ (kpc) & $r(\nu_{12})$ (kpc) \\ 
\midline
0.00 & 105   & 1653 & 804 & $-$  & 33   \\
0.25 & 103.4 & 1541 & 760 & $-$  & 31   \\
0.50 & 92    & 1212 & 593 & 52   & 27.6 \\
0.75 & 60.7  & 997  & 442 & 44.8 & 23   \\
0.90 & 28.4  & 910  & 386 & 32   & 16.3 \\
\bottomline
\end{tabular}
\end{center}
\end{table}

The signal associated to a gravitational collapse can be modeled,
within JBD theory, as a short pulse of amplitude $b\/$, whose value can be
estimated as\cite{shibata}

\begin{equation}
b \simeq 10^{-23}\,\left(\frac{500}{\Omega_{BD}}\right)
  \left(\frac{M_*}{M_\odot}\right)\left(\frac{10\,Mpc}{r}\right)
\label{4.2}
\end{equation}
where $M_*$ is the collapsing mass.

The minimum value of the Fourier transform of the amplitude of the scalar
wave, for a quantum limited detector at unit signal-to-noise ratio, is
given by

\begin{equation}
\left|b(\omega_{nl})\right|_{\min } = \left(
\frac{4\hbar}{Mv_S^2\omega_{nl}K_n}\right)^{1/2}
\label{4.3}
\end{equation}
where $K_n=2H_n$ for the mode with $l=0$ and $K_n=F_n/3$ for the 
mode with $l=2, m=0$.

The duration of the impulse, $\tau \approx 1/f_c$, is much shorter than
the decay time of the $nl\/$ mode, so that the relationship between $b\/$
and $b(\omega_{nl})$ is

\begin{equation}
b \approx \left| b(\omega_{nl})\right| f_c\qquad {\rm at frequency\ \ \ }
  \omega_{nl} = 2\pi f_c
  \label{4.4}
\end{equation}
so that the minimum scalar wave amplitude detectable is

\begin{equation}
\left| b\right|_{\min}\approx\left(
  \frac{4\hbar\omega_{nl}}{\pi^2Mv_S^2K_n}\right)^{1/2}
  \label{4.5}
\end{equation}

Let us now consider a hollow sphere made of molibdenum, for which the speed
of sound is as high as $v_S\/$\,=\,5600 m/sec. For a given detector mass and
diameter, equation (\ref{4.5}) tells us which is the minimum signal detectable
with such detector. For example, a solid sphere of $M=31$ tons and 1.8 metres in
diameter, is sensitive down to $b_{\min}$\,=\,1.5\,$\cdot$\,10$^{-22}$.
Equation (\ref{4.2}) can then be inverted to find which is the maximum
distance at which the source can be identified by the scalar waves it emits.
Taking a reasonable value of $\Omega_{BD}\/$\,=\,600, one finds that
$r(\nu_{10})$\,$\approx$\,0.6 Mpc.

Like before, I report in tables \ref{t.4}, \ref{t.5} and \ref{t.6} the 
sensitivities
of the detector and consequent maximum distance at which the source appears
visible to the device for various values of the thickness parameter
$\varsigma\/$. In table~\ref{t.5} a detector of mass of 31 tons has
been assumed for all thicknesses, and in tables~\ref{t.4}, \ref{t.6} a 
constant outer
diameter of 3 and 1.8 metres has been assumed in all cases.

\begin{table}[ht]
\begin{center}
\footnotesize\rm
\caption{Eigenfrequencies, maximum sensitivities and distances at which
a gravitational collapse can be seen by monitoring the scalar GWs it emits.
Figures correspond to a 31 ton Mb hollow sphere.}
\label{t.5}
\begin{tabular}{lllll}
\topline
$\varsigma $ & $\phi$ (m) & $\nu_{10}$ (Hz) & $|b|_{\min }$ (10$^{-22}$)
& $r(\nu_{10})$ (Mpc) \\ \midline
0.00 & 1.80 & 3338 & 1.5  & 0.6  \\
0.25 & 1.82 & 3027 & 1.65 & 0.5  \\
0.50 & 1.88 & 2304 & 1.79 & 0.46 \\
0.75 & 2.16 & 1650 & 1.63 & 0.51 \\
0.90 & 2.78 & 1170 & 1.39 & 0.6  \\
\bottomline
\end{tabular}
\end{center}
\end{table}

\begin{table}[ht]
\begin{center}
\footnotesize\rm
\caption{Eigenfrequencies, maximum sensitivities and distances at which
a gravitational collapse can be seen by monitoring the scalar GWs it emits.
Figures correspond to a 1.8 metres outer diameter Mb hollow sphere.}
\label{t.6}
\begin{tabular}{lllll}
\topline
$\varsigma $ & $M$ (ton) & $\nu_{10}$ (Hz) & $|b|_{\min }$ (10$^{-22}$)
& $r(\nu_{10})$ (Mpc) \\ 
\midline
0.00 & 31.0  & 3338 & 1.5  & 0.6  \\
0.25 & 30.52 & 3062 & 1.71 & 0.48 \\ 
0.50 & 27.12 & 2407 & 1.95 & 0.42 \\ 
0.75 & 17.92 & 1980 & 2.34 & 0.36 \\ 
0.90 & 8.4   & 1808 & 3.31 & 0.24 \\
\bottomline
\end{tabular}
\end{center}
\end{table}

\section*{Acknowledgments}
I want to thank very much the organizers of the school for their
kind invitation to give these lectures and all of my collaborators
to share their insights on the subject with me.


\newpage
\begin{thebibliography}{99}

\bibitem{ama}{K.S. Thorne in {\it Three hundred years of gravitation}, 
S.W. Hawking and W. Israel editors, Cambridge University Press
(Cambridge, 1987);}

\bibitem{odia}{E. Coccia, G. Pizzella 
and F. Ronga {\it Gravitational Wave Experiments}, 
Proceedings of the First 
Edoardo Amaldi Conference, Frascati 1994, World Scientific Publ. Co.
(Singapore, 1995)}

\bibitem{asto} {P. Astone et al., {\it Phys. Rev.} {\bf D 47} 
(1993) 362.}

\bibitem{joal} {W.W. Johnson, in ref. \cite{odia}.}

\bibitem{abbra} {A. Abramovici et al., {\it Science} {\bf 26} 
(1992) 325;\\
C. Bradaschia et al., {\it Nucl. Instr. Meth.} {\bf A289} (1990) 518.} 

\bibitem{fuc} 
M. Bianchi, E. Coccia, C. N. Colacino, V. Fafone, and
F. Fucito, {\em Class. and Quantum Grav.} {\bf 13}, 2865 (1996);
M. Bianchi, M. Brunetti, E. Coccia, F. Fucito, and
J.A. Lobo, {\em Phys. Rev.} D 57, 4525 (1998); 
M. Brunetti, E. Coccia, V. Fafone and F. Fucito, 
{\em Phys. Rev.} D 59, 044027 (1999);
E. Coccia, F. Fucito, J.A. Lobo and M. Salvino to 
appear in {\em Phys. Rev.}.  

\bibitem{will}
{C.M.~Will, {\it Theory and Experiment in 
Gravitational Physics}, Cambridge University Press 
(Cambridge, 1993).}

\bibitem{elast} {P. Jaerish,{\it J.f. Math.} (Crelle) 
{\bf Bd.88} (1880);\\ 
H. Lamb, in {\it Proceedings of the London Mathematical 
Society} (1882) 
vol. 13; \\
A.E.H. Love, {\it A treatise on the mathematical theory of 
elasticity}, Cambridge University Press, London, 1927.}

\bibitem{lobo} {J.A. Lobo, {\it Phys. Rev.} {\bf D 52} 
(1995) 591.} 

\bibitem{drei} {N.Ashby and J.Dreitlein, {\it Phys. Rev.} 
{\bf D12} (1975)
336.}

\bibitem{jack} {J.D.Jackson, {\it Classical Electrodynamics}, 
John Wiley $\&$ Sons (New York 1975).}

\bibitem{mtw}
{C.W. Misner, K.S. Thorne and J.A. Wheeler, {\it Gravitation},
Freeman (New York, 1973).}

\bibitem{wago} {R.V. Wagoner and H.J. Paik, in 
{\it Experimental Gravitation}, 
Proceeding of the International Symposium held in Pavia 1976 
(Acc. Naz. dei Lincei, Roma, 1977).}

\bibitem{tinto} {S.V. Dhurandhar and M.Tinto, in 
{\it Proceedings of the V
Marcel Grossmann Meeting}, eds. D.G. Blair, M.J. 
Buckingham and R. Ruffini,
World Scientific Publ. Co. (Singapore 1989).}

\bibitem{zhou} {C.Z. Zhou and P.F. Michelson, 
{\it Phys. Rev.} {\bf D 51}
(1995) 2517.}

\bibitem{merko} {W.W. Johnson and S.M. Merkovitz, 
{\it Phys. Rev. Lett.}
{\bf 70} (1993) 2367.}

\bibitem{bd}  P.~Jordan {\it Z. Phys.} {\bf 157}, 112 (1959);  C.~Brans and
R.H.~Dicke, {\it Phys. Rev.} {\bf 124} 925 (1961).

\bibitem{mssr} J.~Miller, M.~Schneider, M.~Soffel and H.~Ruder, 
{\it Astrophys. J. Lett.} {\bf 382}, L101 (1991). 
 
\bibitem{lee} 
D.L.~Lee, {\it Phys. Rev} {\bf D10}, 2374 (1974).

\bibitem{dewey} D.~Dewey, {\it Phys. Rev.} {\bf D36}, 1577 (1987).

\bibitem{cf} E.~Coccia and V.~Fafone, {\it Phys. Letters} {\bf A213}, 16 (1996).

\bibitem{pizz} P.~Astone, G.V.~Pallottino and G.~Pizzella,
{\it Class. Quantum Grav.} {\bf 14}, 2019 (1997).

\bibitem{clark} J.~P.~A.~Clark and D.~M.~Eardley, {\it Astrophys. J.}
{\bf 215}, 311 (1977).

\bibitem{fc} G.~Frossati and E.~Coccia, {\it Cryogenics}
{\bf 34} (ICEC Supplement), 9 (1994).
304 (1994).

\bibitem{ligo} F.J. Raab (LIGO team), in E. Coccia, G.Pizzella, F.Ronga
(eds.), {\em Gravitational Wave Experiments}, Proceedings of the First
Edoardo Amaldi Conference, Frascati 1994 (World Scientific, Singapore, 1995).

\bibitem{virgo} A. Giazotto {\it et al.\/} (VIRGO collaboration), in
E. Coccia, G.Pizzella, F.Ronga (eds.), {\em Gravitational Wave Experiments},
Proceedings of the First Edoardo Amaldi Conference, Frascati 1994 (World
Scientific, Singapore, 1995).

\bibitem{vega-98} E. Coccia, V. Fafone, G. Frossati, J. A. Lobo, 
and J. A. Ortega, {\it Phys. Rev.} D 57, 2051 (1998).


\bibitem{wein-72} S. Weinberg, {\it Gravitation and Cosmology\/}, Wiley
\& sons, New York 1972.

\bibitem{shibata}
M.~Shibata, K.~Nakao and T.~Nakamura,
{\em Phys. Rev.} D 50, 7304 (1994); M.~Saijo, H.~Shinkai and K.~Maeda,
{\em Phys. Rev.} D 56, 785 (1997).
\end{thebibliography}
\end{document}